\documentclass[aps,prl,twocolumn,groupedaddress,showpacs]{revtex4}

\usepackage{graphicx}
\begin{document}

%\preprint{}

\title{Single-shot measurement of quantum optical phase}


\author{K. L. Pregnell and D. T. Pegg}
%\thanks{}
\affiliation{School of Science, Griffith University, Nathan, Brisbane,
4111, Australia}


\date{\today}

\begin{abstract}
Although the Canonical phase of light, which is defined as the complement of photon number, has been described
theoretically by a variety of distinct approaches, there have been no methods proposed for its measurement. Indeed
doubts have been expressed about whether or not it is measurable. Here we show how it is possible, at least in
principle, to perform a single-shot measurement of Canonical phase using beam splitters, mirrors, phase shifters and
photodetectors.
\end{abstract}

%\keywords{}

\maketitle Quantum-limited phase measurements of the optical field have important applications in precision
measurements of small distances in interferometry and in the emerging field of quantum communication, where there is
the possibility of encoding information in the phase of light pulses. Much work has been done in attempting to
understand the quantum nature of phase.  Some approaches have been motivated by the aim of expressing phase as the
complement of photon number \cite{Leon}. Examples of these approaches include the probability operator measure approach
\cite{Helstrom,SS}, a formalism in which the Hilbert space is doubled \cite{Newton}, a limiting approach based on a
finite Hilbert space \cite{PB,tute} and a more general axiomatic approach \cite{Leon}.  Although these approaches are
quite distinct, they all lead to the same phase probability distribution for a field in state $|\psi\rangle$ as a
function of the phase angle $\theta$ \cite{Leon}:
\begin{equation}
P(\theta) =\frac{1}{2\pi}|\sum_{n=0}^\infty \langle\psi|n\rangle
\exp (in\theta)|^{2}  \label{0}
\end{equation}
where $|n\rangle$ is a photon number state. Leonhardt {\em et al.} \cite{Leon} have called this common distribution the
``canonical'' phase distribution to indicate a quantity that is the canonical conjugate, or complement, of photon
number. This distribution is shifted uniformly when a phase shifter is applied to the field and is not changed by a
photon number shift \cite{Leon}. We adopt this definition here and use the term Canonical phase to denote the quantity
whose distribution is given by (\ref{0}).

Much less progress has been made on ways to measure Canonical phase. Homodyne techniques can be used to measure
phase-like properties of light but are not measurements of Canonical phase. It is possible in principle to measure the
Canonical phase distribution by a series of experiments on a reproducible state of light \cite{BPprl} but there has
been no known way of performing a single-shot measurement. Indeed it is thought that this might be impossible
\cite{Wiseman}. Even leaving aside the practical issues, the concept that a particular fundamental quantum observable
may not be measurable, even in principle, has interesting general conceptual ramifications for quantum mechanics. A
different approach to the phase problem, which avoids difficulties in finding a way to measure Canonical phase, is to
define phase operationally in terms of observables that can be measured \cite {Leon}. The best known of these
operational phase approaches is that of Noh {\em et al.} \cite {Noh1, Noh2}. Although the experiments to measure this
operational phase produce excellent results, they were not designed to measure Canonical phase as defined here and, as
shown by the the measured phase distribution \cite{Noh2}, they do not measure Canonical phase. In this paper we show
how, despite these past difficulties, it is indeed possible, at least in principle, to perform a single-shot
measurement of Canonical phase in the same sense that the experiments of Noh {\em et al.} are single-shot measurements
of their operational phase.

A single-shot measurement of a quantum observable must not only yield one of
the eigenvalues of the observable, but repeating the measurement many times
on systems in identical states should result in a probability
distribution appropriate to
that state. If the spectrum of eigenvalues is discrete, the probabilities
of the results can be easily obtained from the experimental statistics.
Where the spectrum is
continuous, the probability density is obtainable by dividing the eigenvalue
range into a number of small bins and finding the number of results in each
bin. As the number of experiments needed to obtain measurable
probabilities increases as the reciprocal of the bin size, a practical
experiment will require a non-zero bin size and will produce a histogram
rather than a smooth curve.

Although the experiments of Noh {\em et al.} \cite{Noh1, Noh2} were not designed to measure Canonical phase, it is
helpful to be guided by their approach. In addition to their results being measured and plotted as a histogram, some of
the experimental data are discarded, specifically photon count outcomes that lead to an indeterminacy of the type zero
divided by zero in their definitions of the cosine and sine of the phase \cite{Noh1}. The particular experiment that
yields such an outcome is ignored and its results are not included in the statistics.

Concerning the discarding of some results, we note in general that the well-known expression for the probability that a
von Neumann measurement on a pure state $|\psi \rangle $ yields a result $q$ is $\langle \psi |q\rangle \langle q|\psi
\rangle $, where $|q\rangle $ is the eigenstate of the measurement operator corresponding to eigenvalue $q$. If the
state to be measured is a mixed state with a density operator $\hat{\rho }$ the expression becomes Tr($\hat{\rho }
\widehat{\Pi }_q)$ where $\widehat{\Pi }_q$ = $|q\rangle \langle q|$. The operator $\widehat{\Pi }_q$ is a particular
case of an element of a probability operator measure (POM) \cite{Helstrom}. The sum of all the elements of a POM is the
unit operator and the expression Tr($\hat{\rho }\widehat{\Pi }_q)$ for the probability is based on the premise that all
possible outcomes of the measurement are retained for the statistics. If some of the possible outcomes of an experiment
are discarded, the probability of a particular result calculated from the final statistics is given by the normalized
expression Tr($\hat{\rho }\widehat{\Pi }_q)/\sum_p$Tr($\hat{\rho } \widehat{\Pi }_p)$, where the sum is over all the
elements of the POM corresponding to outcomes of the measurement that are retained.

We seek now to approximate the continuous distribution (\ref{0}) by a
histogram representing the probability distribution for a discrete observable
$\theta_{m}$ such that when the separation $\delta\theta$ of consecutive
values of $\theta_{m}$ tends to zero the continuous distribution is
regained.  A way to do this is first to define a state
\begin{equation}
|\theta_m\rangle =\frac 1{(N+1)^{1/2}}\sum_{n=0}^N\exp (in\theta_m)|n\rangle.
\label{1}
\end{equation}
There are $N+1$ orthogonal states $|\theta_{m}\rangle$ corresponding
to $N+1$ values $\theta_{m}=m\delta\theta$ with $\delta\theta =
2\pi/(N+1)$ and $m=0,1,\dots N$.  This range for $m$ ensures that
$\theta_{m}$ takes values between $0$ and $2\pi$. Then, if we can find
a measurement technique that yields the result $\theta_{m}$ with a
probability of $|\langle \psi |\theta _m\rangle |^2$, the resulting
histogram will approximate a continuous distribution with a
probability density of $|\langle \psi |\theta _m\rangle
|^2/\delta\theta$. It follows that, as we let $N$ tend to infinity,
there will exist a value of $\theta_{m}$ as close as we like to any
given value of $\theta$ with a probability density approaching
$P(\theta)$ given by (\ref{0}). If we keep $N$ finite so that we can
perform an experiment with a finite number of outcomes, then the value
of $N$ must to be sufficiently large to give the resolution
$\delta\theta$ of
phase angle required and also for $|\psi \rangle$ to be well approximated by
$\sum_{n} \langle n|\psi \rangle |n\rangle$ where the sum is from
$n = 0$ to $N$. The latter condition ensures that the terms with
coefficients $\langle n|\psi\rangle$ for $n > N$ have little effect
on the probability $|\langle \psi |\theta _m\rangle |^2$. As we shall be
interested mainly in weak optical fields in the quantum regime with mean
photon
numbers of the order of unity, the maximum phase resolution
$\delta\theta$ desired will usually be the
determining factor in the choice of $N$.

When $N$ is finite, the
states $|\theta _m\rangle $ do not span the whole Hilbert space, so the
projectors $|\theta _m\rangle \langle \theta _m|$ will not sum to the unit
operator $\hat{1}$. Thus these projectors by themselves do not form
the elements of a POM. To complete the POM we need to include an element $
\hat{1}-\sum_m|\theta _m\rangle \langle \theta _m|$. If we discard the
outcome associated with this element, that is, treat an experiment with this
outcome  as an
unsuccessful attempt at a measurement in a similar way that
Noh {\it et al.} \cite{Noh1} treated experiments with indeterminate
outcomes, then the probability that the
outcome of a measurement is the phase angle $\theta _m$ is given by
\begin{equation}
Pr(\theta _m)=\frac{\text {Tr}(\hat{\rho }|\theta _m\rangle \langle \theta
_m|)}
{\sum_p\text {Tr}(\hat{\rho }|\theta _p\rangle \langle \theta _p|)},  \label{2}
\end{equation}
where $p=0,1\ldots\,N$. We now require a single-shot measuring device that
will reproduce this probability in repeated experiments.

\begin{figure}
\includegraphics{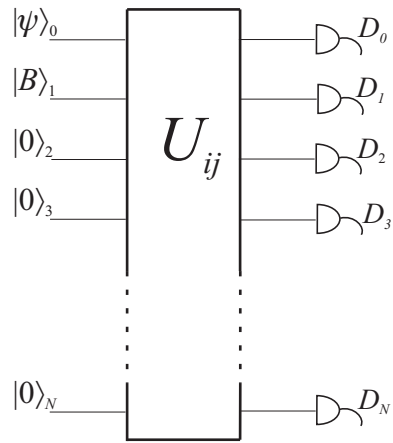}
\caption{Multi-port device for measuring phase. The input and output
modes are labelled $0, 1, \ldots N$ from the top. In input mode $0$
is the field in state $|\psi\rangle_{0}$ to be measured and in input mode
$1$ is the reference field in state $|B\rangle_{1}$. Vacuum states form
the other inputs. There is a photodetector in each output mode. If
all the photodetectors register one count except the detector $D_{m}$
in output
mode $m$, which registers no counts, then the detector array acts as
a digital pointer mechanism indicating a measured phase angle of
$\theta_{m}$.\label{Fig1}}
\end{figure}

As measurements will be performed ultimately by photodetectors, we seek
an optical device that transforms input fields in such a way
that photon number measurements at the output ports can be converted
to phase measurements.
We examine the case of a multi-port device with $N+1$ input modes and $N+1$
output modes as depicted schematically in Fig. \ref{Fig1}. As phase is not
an absolute
quantity, that is it can only be measured in relation to some reference
state, we shall need the field in state $|\psi \rangle _0$ that is
to be measured to be in one input and a reference
field in state $|B\rangle _{1\text{ }}$ to be in another input. We let there be
vacuum states $|0\rangle _i$ with $i=2,3\ldots N$ in the other input modes.
We let the device be such that the input states are transformed by a unitary
transformation $\widehat{R}$. We let $N+1$ photodetectors that can
distinguish between
zero photons, one photon and more than one photon be in the output modes.

Consider the case where one photon is detected in each output mode except
for mode $m$, in which zero photons are detected. The amplitude for this
detection event is
\begin{equation}
_m\langle 0|\left (\prod_{j\neq m}\,_j\langle 1|\right )\widehat{R}
\left (\prod_{i=2}^N|0\rangle
_i\right )|B\rangle _1|\psi \rangle _0\,  \label{3}
\end{equation}
where the first product is over values of $j$ from $0$ to $N$ excluding the
value $m$. We require the transformation and the reference state to be such
that this amplitude is proportional to $_0\langle \theta _m|\psi \rangle _0$
. We write the photon creation operators acting on the mode $i$ as $\hat{
a}_i^{\dagger}$. Writing $|1\rangle _j$ as $\hat{a}_j^{\dagger
}|0\rangle _j$ we see that we require $|\theta _m\rangle _0$ to be
proportional to
\begin{equation}
_1\langle B|\left (\prod_{i=2}^N\,_i\langle 0|\right )\left (\prod_{j\neq m}
\widehat{R}^{\dagger}\hat{a}_j^{\dagger }\widehat{R}\right )
\widehat{R}^{\dagger }\,\left (\prod_{k=0}^{N}|0\rangle
_k\right )  \label{3a}
\end{equation}
 We rewrite the unitary transformation in the form
\begin{equation}
\widehat{R}^{\dagger }\hat{a}_j^{\dagger }\widehat{R}=\sum_{i=0}^NU_{ij}
\hat{a}_i^{\dagger }  \label{4}
\end{equation}
where $U_{ij}$ are the elements of a unitary matrix. Reck {\em et al.} \cite
{Reck} have shown how it is possible to construct a multi-port device from
mirrors, beam splitters and phase shifters that will transform the input
modes into the output modes in accord with any $(N+1)\times (N+1)$ unitary
matrix. We choose such a device for which the
associated unitary matrix is
\begin{equation}
U_{ij}=\frac{\omega ^{ij}}{\sqrt{N+1}}  \label{5}
\end{equation}
where $\omega =$ $\exp[-i2\pi /(N+1)]$ that is, a $(N+1)$th root of unity.
Such a device will transform the combined vacuum state to the
combined vacuum state so we can delete the $\widehat R^{\dagger}$ on the
right of expression (\ref{3a}).
Substituting (\ref{5}) into (\ref{4}) gives eventually
\begin{eqnarray}
\left (\prod_{i=2}^N\,_i\langle 0|\right )\left (\prod_{j\neq m}
\widehat{R}^{\dagger}\hat{a}_j^{\dagger }\widehat{R}\right )
\left (\prod_{k=0}^{N}|0\rangle_k\right )= \nonumber
\\
\kappa_{1}\left [\prod_{j\neq m}(\hat{a}_0^{\dagger }+\omega ^j\hat{a}
_1^{\dagger })\right ] |0\rangle _0|0\rangle _1  \label{6}
\end{eqnarray}
where $\kappa_{1} =(N+1)^{-N/2}$.

To evaluate (\ref{6}) we divide both sides of the identity
\begin{equation}
X^{N+1}+(-1)^N=(X+1)(X+\omega )(X+\omega ^2)\ldots (X+\omega ^N)  \label{7}
\end{equation}
by $X+\omega ^m$ to give, after some rearrangement and application of the
relation $\omega ^{m(N+1)}=1$,
\begin{equation}
\prod_{j\neq m}(X+\omega ^j)=(-1)^N\omega ^{mN}\frac{1-(-X\omega
^{-m})^{N+1}
}{1-(-X\omega ^{-m})}\text{ .}  \label{8}
\end{equation}
The last factor is the sum of a geometric progression. Expanding this and
substituting $X=x/y$ gives eventually the identity
\begin{equation}
\prod_{j\neq m}(x+\omega ^jy)=\sum_{n=0}^N x^n(-\omega
^my)^{N-n}\text{
.}  \label{9}
\end{equation}

We now expand $|B\rangle _1$ in terms of photon number states as
\begin{equation}
|B\rangle _1=\sum_{n=0}^Nb_n|n\rangle_{1} \label{10}
\end{equation}
and put $x=a_0^{\dagger }$ and $y=a_1^{\dagger }$ in (\ref{9}). Then from (\ref
{6}) we find that (\ref{3a}) becomes
\begin{eqnarray}
_1\langle B|\left (\prod_{i=2}^N\,_i\langle 0|\right )\left (\prod_{j\neq m}
\widehat{R}^{\dagger}\hat{a}_j^{\dagger }\widehat{R}\right )
\widehat{R}^{\dagger }\,\left (\prod_{k=0}^{N}|0\rangle
_k\right ) =\nonumber
\\ \kappa_{2}\sum_{n=0}^{N}(-1)^{N-n}{N\choose
n}^{-1/2}\omega^{-nm}b_{N-n}^{*}|n\rangle_{0}
\label{11}
\end{eqnarray}
where $\kappa_{2}= \kappa_{1}\omega^{-m}(N!)^{1/2}$. We see
then that, if we let $|B\rangle_{1}$ be the binomial state
\begin{equation}
   |B\rangle_{1}=2^{-N/2}\sum_{n=0}^{N}(-1)^{n}{N\choose
   n}^{1/2}|n\rangle_{1},
   \label{12}
 \end{equation}
then expression (\ref{11}) is proportional to $\sum_{n}\omega^{-nm}|n\rangle_{0}$, that is, to
$|\theta_{m}\rangle_{0}$. Thus the amplitude for the event that zero photons are detected in output mode $m$ and one
photon is detected in all the other output modes will be proportional to $_{0}\langle \theta_{m}|\psi\rangle_{0}$. The
probability that the outcome of a measurement is this event, given that only outcomes associated with the $(N+1)$
events of this type are recorded in the statistics, will be given by (\ref {2}), where we note that the proportionality
constant $\kappa_{2}$ will cancel from this expression. Thus the measurement event that zero photons are detected in
output mode {\em m} and one photon is detected in all the other output modes can be taken as the event that the result
of the measurement of the phase angle is $\theta_{m}$. Thus the photodetector with zero photocounts, when all other
photodetectors have registered one photocount, can be regarded as a digital pointer to the value of the measured phase
angle.

We have shown, therefore, that it is indeed possible in principle to conduct a single-shot measurement of Canonical
phase to within any given non-zero error, however small. This error is of the order $2\pi/(N+1)$ and will determine the
value of $N$ chosen.

\begin{figure}
\includegraphics{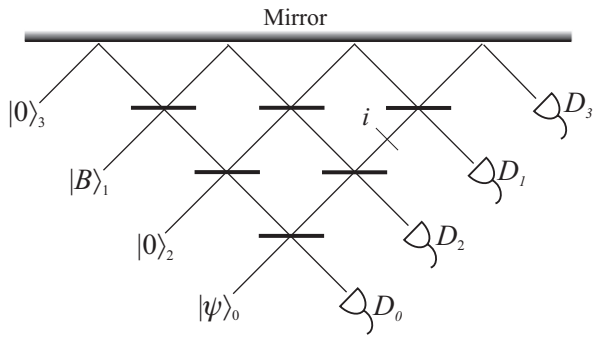}
\caption{Triangular array for $N+1 = 4$. The outside beam splitters in the top row are 50:50, the middle one is fully
reflecting. The second row beam splitters are $\frac{2}{3}$-transmitting and the bottom one is
$\frac{3}{4}$-transmitting. The phase-shifter $i$ produces a $\pi/2$ phase shift. \label{Fig2}}
\end{figure}

While the aim of this paper is to establish how Canonical phase can be measured in principle, it is worth briefly
considering some practical issues. Although we have specified that the photodetectors need only be capable of
distinguishing among zero, one and more than one photons, reflecting the realistic case, there are other imperfections
such as inefficiency. These will give rise to errors in the phase measurement, just as they will cause errors in a
single-shot photon number measurement. In practice, there is no point in choosing the phase resolution $\delta\theta$
much smaller than the expected error due to photodetector inefficiencies, thus there is nothing lost in practice in
keeping $N$ finite. A requirement for the measuring procedure is the availability of a binomial state. Such states have
been studied for some time \cite{binom} but their generation has not yet been achieved. In practice, however, we are
usually interested in measuring weak fields in the quantum regime with mean photon numbers around unity \cite{Noh1,
Noh2} and even substantially less \cite{Torg}.  Only the first few coefficients of $|n\rangle_{0}$ in (\ref{11}) will
be important for such weak fields. Also, it is not difficult to show that the reference state need not be truncated at
$n = N$, as indicated in (\ref{10}), as coefficients $b_{n}$ with $n > N$ will not appear in (\ref{11}). Thus we need
only prepare a reference state with a small number of its photon number state coefficients proportional to the
appropriate binomial coefficients. Additionally, of course, in a practical experiment we are forced to tolerate some
inaccuracy due to photodetector errors, so it will not be necessary for the reference state coefficients to be exactly
proportional to the corresponding binomial state coefficients. These factors give some latitude in the preparation of
the reference state. The muti-port device depicted in Fig. 1 can be constructed in a variety of ways. Reck {\em et al.}
\cite{Reck} provide an algorithm for constructing a triangular array of beam splitters to realize any unitary
transformation matrix. Fig. 2 shows, for example, such a device that, with suitable phase shifters in the output modes,
will realize the transformation $U_{ij}$ in (\ref{5}) with $N+1 = 4$. Because we are detecting photons, however, these
output phase shifters are not actually necessary and are thus not shown. The number of beam splitters needed for a
general triangular array increases quadratically with $N$. Fortunately, however, our required matrix (\ref{5})
represents a discrete Fourier transformation and we only require two of the input ports to have input fields that are
not in the vacuum state. The device of T\"{o}rm\"{a} and Jex \cite{Jex} is ideally suited for these specific
requirements. This device, which has an even number of input and output ports, is pictured in Ref. \cite{Jex}. It
consists of just $(N+1)/2$ ordinary 50:50 beam splitters and two plate beam splitters. The latter are available in the
form of glass plates with modulated transmittivity along the direction of the incoming beam propagation \cite{Jex}.

As a large fraction of the raw data in this procedure is discarded, an interesting question arises as to whether or not
there is a relationship between the method of this paper and a limiting case of the operational phase measurements of
Torgerson and Mandel \cite{Torg2} where it is found that the distribution becomes sharper as more data are discarded.
While preliminary analysis indicates that there is not, this will be discussed in more detail elsewhere.

In conclusion we have shown that it is possible in principle to perform a single-shot measurement of the Canonical
phase in the same sense that the experiments of Noh {\it et al.} are single-shot measurements of operational phase. The
technique relies on generating a reference state with some number state coefficients proportional to those of a
binomial state.

\begin{acknowledgments}
D. T. P. thanks the Australian Research Council for funding.
\end{acknowledgments}


\end{document}